\documentclass[t wocolumn,showpacs,preprintnumbers,amsmath,amssymb]{revtex4}
\usepackage{graphicx}
\usepackage{dcolumn}
\usepackage{bm}

\newcommand{\bef}{\begin{figure}}
\newcommand{\eef}{\end{figure}}

\newcommand{\be}{\begin{equation}}
\newcommand{\ee}{\end{equation}}
\newcommand{\bea}{\begin{eqnarray}}
\newcommand{\eea}{\end{eqnarray}}
\begin{document}

\title{Baseline measures for net-proton distributions in high energy heavy-ion collisions}
\author{P.~K.~Netrakanti$^1$, X.~F.~Luo$^2$, D.~K.~Mishra$^1$, B.~Mohanty$^3$,
  A.~Mohanty$^1$ and N.~Xu$^{2,4}$}
\affiliation{ $^1$Nuclear Physics Division, Bhabha Atomic Research
  Center, Mumbai 400094, India, 
$^2$Key Laboratory of the Ministry of Education of China, Central
China Normal University, Wuhan, 430079, China, 
$^3$School of Physical Sciences, National Institute of Science Education and Research,
Bhubaneswar-751005, India, and $^4$Nuclear Science Division, Lawrence Berkeley National Laboratory, Berkeley, CA 94720, USA}

\date{\today}
\begin{abstract}
We report a systematic comparison of the recently measured cumulants
of the net-proton distributions for 0-5\% central Au+Au collisions
in the first phase of the Beam Energy Scan (BES) Program at the Relativistic
Heavy Collider facility to various kinds of possible baseline measures. These
baseline measures correspond to assuming that the proton and anti-proton
distributions, follow Poisson statistics, Binomial statistics,
obtained from a transport model calculation and from a
hadron resonance gas model. The higher order cumulant net-proton data
corresponding to the center of mass energies ($\sqrt{s_{NN}}$) of 19.6
and 27 GeV  are observed to deviate from all the baseline measures studied. The
deviations are predominantly due to the difference in shape of the proton
distributions between data and those obtained in the baseline measures. 
We also present a detailed study on the relevance of the independent
production approach as a baseline for comparison with the measurements  
at various beam energies. Our studies points to the need for a proper
comparison of the experimental measurements to QCD calculations in
order to extract the exact physics process that leads to deviation of
the data from the baselines presented. 
\end{abstract}
\maketitle

\section{Introduction}

The STAR experiment at the Relativistic Heavy-Ion Collider facility
has recently reported interesting results on the shape of the
net-proton distributions at various collision energies \cite{Adamczyk:2013dal}. These
measurements are carried out as a part of the first phase of the beam
energy scan program to look for the signatures of the possible critical
point (CP) in the phase diagram for a system undertaking strong
interactions. The shape of the net-proton distributions (a proxy for
net-baryon distributions) is quantified in terms of the cumulants of
the distribution. Measurements upto the fourth order cumulants
($C_{n}$, $n$ = 1,2,3, and 4) has been reported as a function of the
colliding beam energy. Varying the beam energy of the collision also
varies the baryon chemical potential of the system. Thereby allowing
experimentally to scan the temperature versus baryon chemical
potential phase diagram of strong interactions. The
STAR experiment has reported an intriguing dependence of the cumulant
ratios $C_{3}/C_{2}$ and $C_{4}/C_{2}$ as a function of beam energy.
The beam energy dependence appears to be non-monotonic in nature. 
However the experiment also reports that the energy dependence is
observed to be consistent with expectation from an approach based on
the independent production of proton and anti-protons in the collisions \cite{Adamczyk:2013dal}.

In this paper we first establish that at the lower colliding energies the
beam energy dependence of
the net-proton cumulant ratios are dominantly due to the corresponding proton
distributions.  Very much similar to recently reported, non-monotonic
beam energy dependence of the slope of the net-proton directed flow at
midrapidity being dominantly due to the corresponding contribution from
the measured proton directed flow~\cite{Adamczyk:2014ipa}. We emphasize the need to have a proper baseline for
appropriate interpretation of the cumulant measurements and argue that the
comparison to independent production approach needs to be done with
extreme caution. We demonstrate through our study that the
applicability of the independent production approach at lower beam
energies where the anti-proton production is very small is
questionable. Further, we have argued that the agreement at the higher
beam energies in-spite of significant correlated production of proton
and anti-proton ($p/\bar{p}$ $\sim$ 0.77, for 0-5\% central Au+Au
collisions at $\sqrt{s_{NN}}$ = 200 GeV~\cite{Abelev:2008ab}) in the collisions could be a
coincidence due to the acceptance in which measurements have been
carried out. In addition we  point out the role
of particle production mechanism and baryon number conservation to
such approaches. We have also carried out a very systematic
comparison of the four measured cumulants to a variety of
baselines measures. These include, expectations if proton and
anti-proton distributions are Poisson, Binomial, those obtained from a
transport model and from a hadron resonance gas (HRG) Model. All these variety
of baselines indicate the higher order cumulant data deviates from them
for central 0-5\% Au+Au collisions at $\sqrt{s_{NN}}$ = 19.6 and 27 GeV. 

The paper is organized as follows. Next section deals with the
experimental data and sets up the physics problem addressed in the
paper. Section III compares the cumulants measured in the experiment
to the four different baseline measures. Section IV discusses in detail the
independent production approach and finally in section V we summarize our
findings.

\section{Experimental data}
\bef
\begin{center}
\includegraphics[scale=0.40]{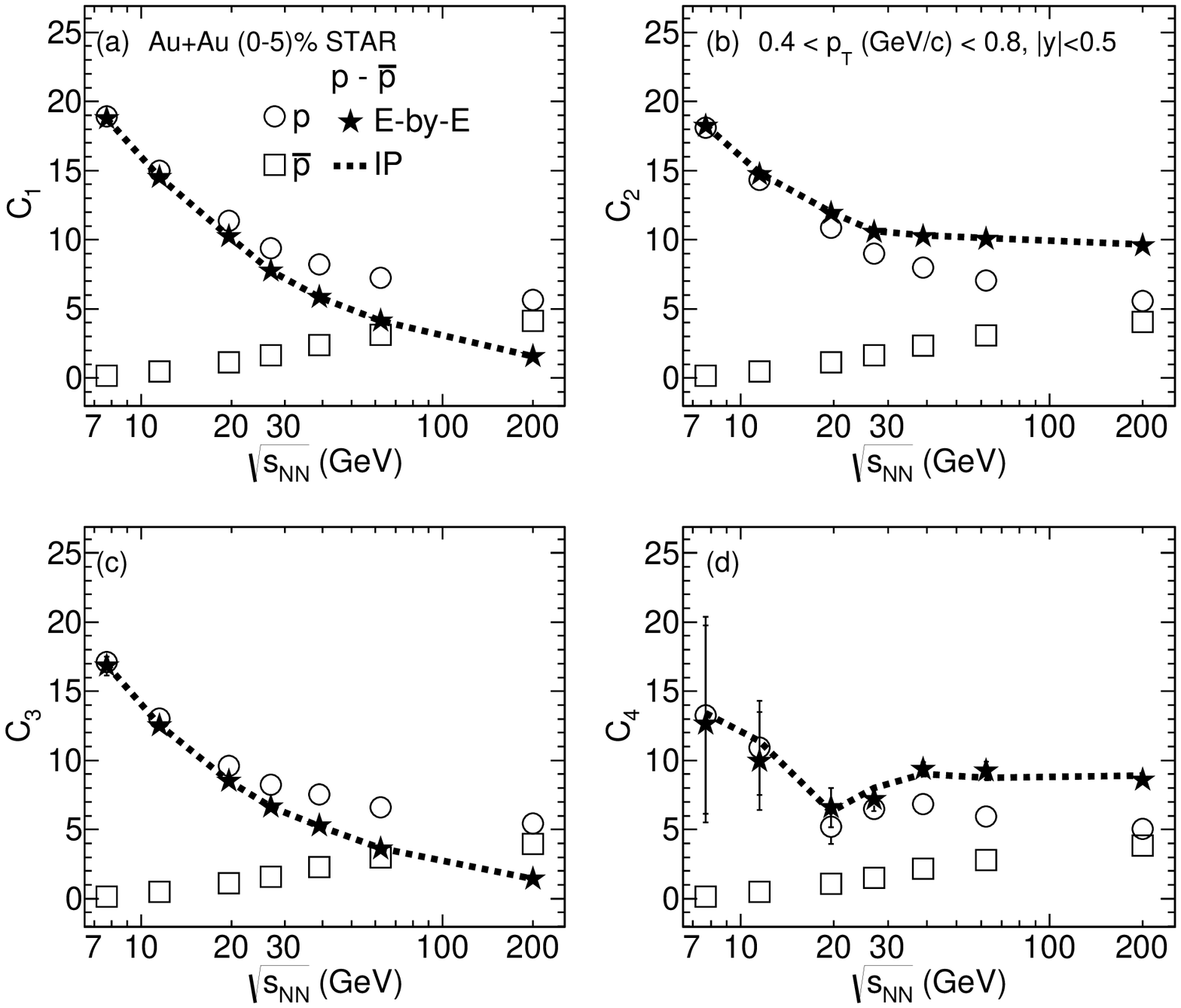}
\caption{Cumulants of proton, anti-proton and net-proton multiplicity
  distribution at midrapidity in 0-5\% central Au+Au collisions at RHIC BES program~\cite{Adamczyk:2013dal}.
Also shown are the expectations from independent production model.}
\label{fig1}
\end{center}
\eef
Figure~\ref{fig1} shows the four cumulants of the measured proton,
anti-proton and net-proton distributions at midrapidity in 0-5\%
central Au+Au collisions as a function of center of mass energy.  The
cumulants of proton distribution in general decreases with increase in beam energy whereas
those for the anti-protons increases. This trend is consistent with
the picture of evolution of baryon stopping and transparency with beam
energy. While the cumulants $C_1$ to $C_3$ seems to show a monotonic
trend with $\sqrt{s_{NN}}$, the $C_{4}$ shows a non-monotonic
behaviour with beam energy. Higher statistics measurements of $C_{4}$ at
$\sqrt{s_{NN}}$ = 11.5 and 7.7 GeV in BES phase-II will make the trend clearer.  
Figure~\ref{fig1} also shows that the cumulants of net-proton
distribution are
dominated by contribution from the corresponding cumulants of proton 
distribution below $\sqrt{s_{NN}}$ = 39 GeV. 

The experimental data on cumulants of net-proton is compared to
results from an independent production (IP) model. In the IP model the
measured cumulants of proton and anti-proton distributions are taken
and corresponding cumulants for net-proton distribution constructed
assuming that proton and anti-proton productions are independent of
each other. They are expected to have errors of the same order as data
by construction. For the IP, the various order ($n$ = 1,2,3 and 4)
net-proton cumulants are given as $C_{n}$ = $C_{n}^{p}$ + $(-1)^{n}$
$C_{n}^{\bar{p}}$, where $C_{n}^{p}$ and $C_{n}^{\bar{p}}$ are the  
cumulants of the measured proton and anti-proton distributions.
IP results seem to explain the experimentally measured 
net-proton cumulants for all beam energies. However, for $\sqrt{s_{NN}}$ $<$ 39
GeV, the IP model applicability is questionable as the net-proton
cumulants are dominated by contributions from protons only. Hence
due to the absence of significant anti-proton production, the
basic assumption in IP model that proton and anti-proton productions
are independent is not applicable and proton distributions solely
reflect the net-proton distributions. At the same time it is known
from the measured $\bar{p}/p$ ratio ($\sim$ 0.77 at midrapidity in
Au+Au collisions at $\sqrt{s_{NN}}$ = 200 GeV~\cite{Abelev:2008ab}) there is a significant level of
correlation between proton and anti-proton production. Hence the
agreement of IP model results at the higher side of the beam energies studied is
also intriguing. Further, the role of baryon number conservation on
such an approach has to be understood properly. Understanding of the IP
model is attempted in the discussions using various models in section
IV. All these also necessitates a discussion on other proper baselines for the 
experimental data, this is carried out in the section below.

\section{Baseline for cumulants of net-proton distribution}

\subsection{Poisson}
\bef
\begin{center}
\includegraphics[scale=0.40]{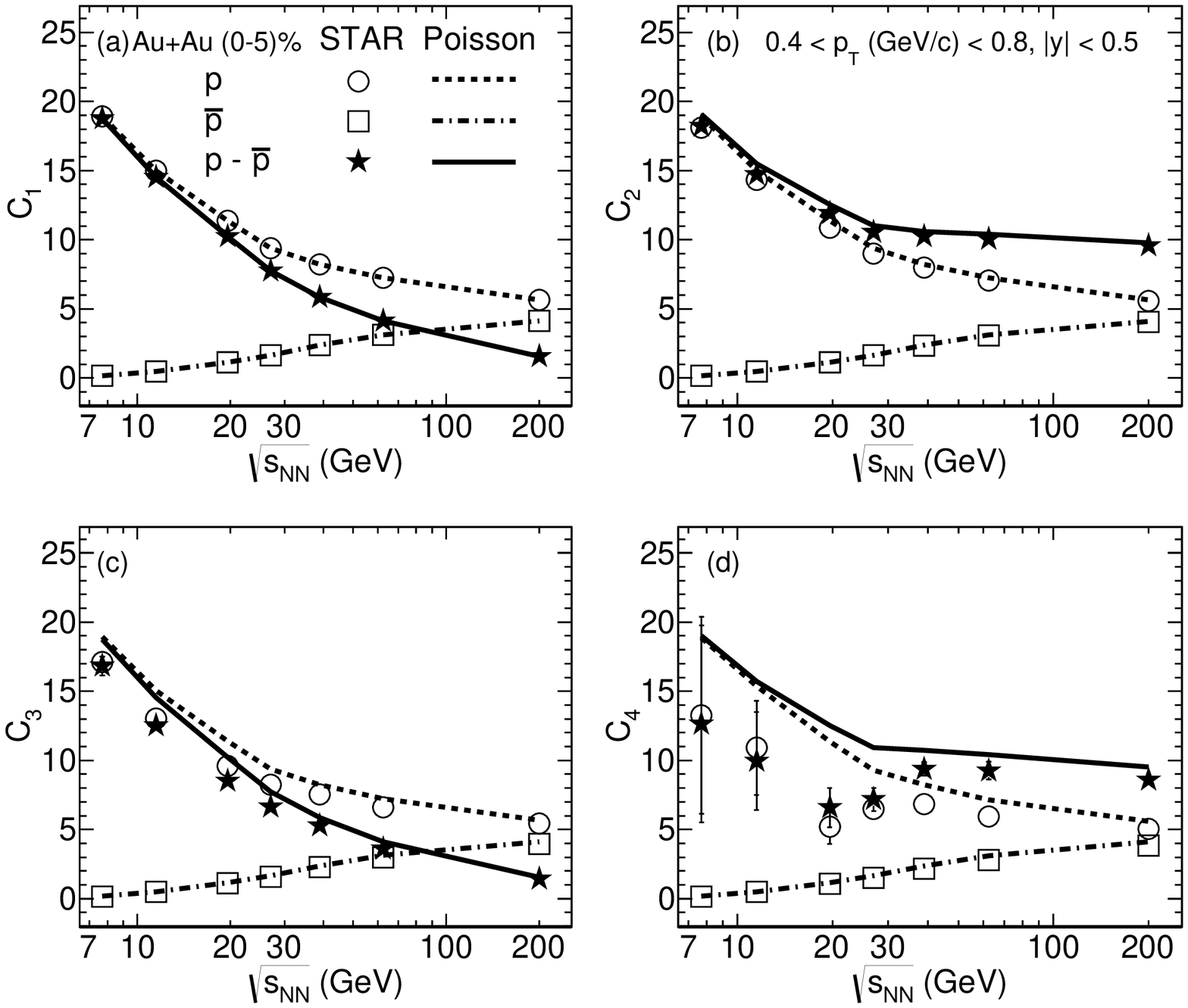}
\caption{Cumulants of proton, anti-proton and net-proton multiplicity
  distribution at midrapidity in 0-5\% central Au+Au collisions at RHIC BES program.
Data are compared to Poisson expectations.}
\label{fig2}
\end{center}
\eef

Poisson distribution represents a statistically random expectation of
the observable. Taking the measured mean value of the number of
protons and anti-protons produced in the heavy-ion collisions one can
construct the corresponding net-proton distribution. The proton and
anti-proton production are assumed to be independent. The resultant
distribution is called the Skellam distribution. Figure~\ref{fig2} shows the comparison of the measured
cumulants for proton, anti-proton and net-proton distributions to
Poisson expectation. As the measured $C_{1}$ is used to construct the
Poisson expectation the agreement is by construction. However
Figure~\ref{fig2}  shows as the order of cumulant increases we find
deviations of the data from the Poisson expectation for protons and
net-protons increases for $\sqrt{s_{NN}}$ $<$ 62.4 GeV. The anti-proton
cumulants are reasonably well explained by the Poisson expectation.
Hence we conclude that data shows deviation from what is expected from
a fully uncorrelated and random statistical process for the proton
distributions and hence the same is reflected for the net-proton distributions for
$\sqrt{s_{NN}}$ $<$ 62.4 GeV.

\subsection{Binomial}
\bef
\begin{center}
\includegraphics[scale=0.40]{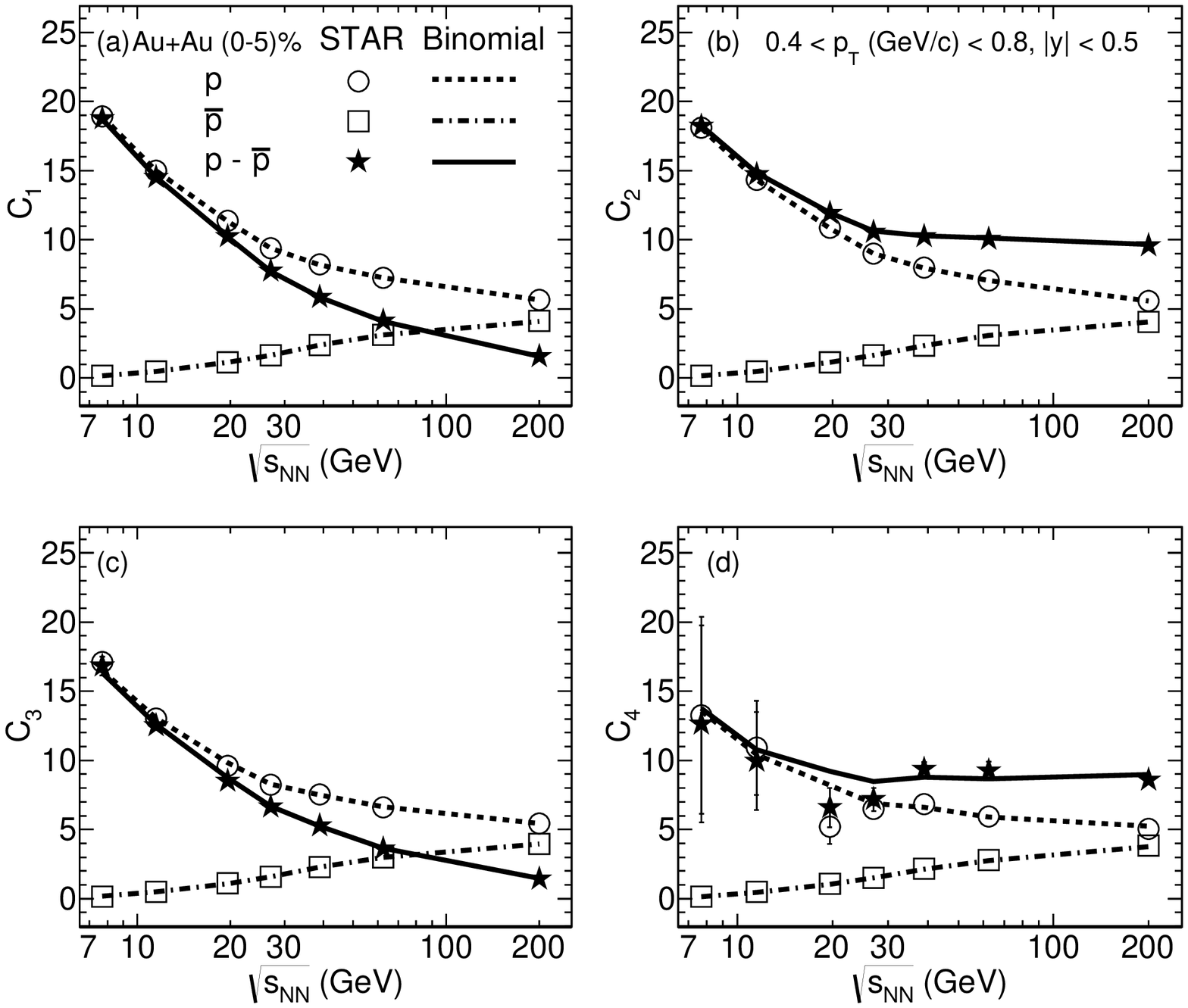}
\caption{Cumulants of proton, anti-proton and net-proton multiplicity
  distribution at midrapidity in 0-5\% central Au+Au collisions at RHIC BES program.
Data are compared to Binomial expectations.}
\label{fig3}
\end{center}
\eef

Figure~\ref{fig3} shows the comparison of the measured
cumulants for proton, anti-proton and net-proton 0-5\% central
collision distributions to Binomial expectation. To construct the binomial distributions, the
measured $C_{1}$ and $C_{2}$ values of the proton and anti-proton
distributions are taken. To obtain the corresponding
distribution for net-proton, it is assumed that the binomially
distributed proton and anti-proton are produced independently. Hence
we see good agreement between data and binomial expectation for
$C_{1}$ and $C_{2}$.  Deviations are observed only for $C_{4}$ of the
measured  proton and hence net-proton distributions for
$\sqrt{s_{NN}}$ = 19.6 and 27 GeV. The cumulants of
the anti-proton distribution like in the above case of Poisson
expectation seems to also follow reasonably well the binomial
expectation for all beam energies measured.  Hence we conclude that
the highest order measured
cumulant for proton and net-proton distribution shows deviation from
binomial expectation for $\sqrt{s_{NN}}$ = 19.6 and 27 GeV. The
difference  between the Poisson expectation
and binomial expectation comparison is that $C_{3}$ is reasonably well
explained by binomial expectation unlike for the Poisson case. In
turn for construction of binomial distribution we make use of both
measured $C_{1}$ and $C_{2}$ of proton and anti-proton distributions
while only $C_{1}$ is used to construct the corresponding distribution
following Poisson statistics. A comparison of HIJING simulation results of 
$C_3/C_2$ and $C_4/C_2$ with the binomial and negative binomial distribution expectation 
are discussed in ref ~\cite{Westfall:2013plb}.

\subsection{Transport Model}
\bef
\begin{center}
\includegraphics[scale=0.40]{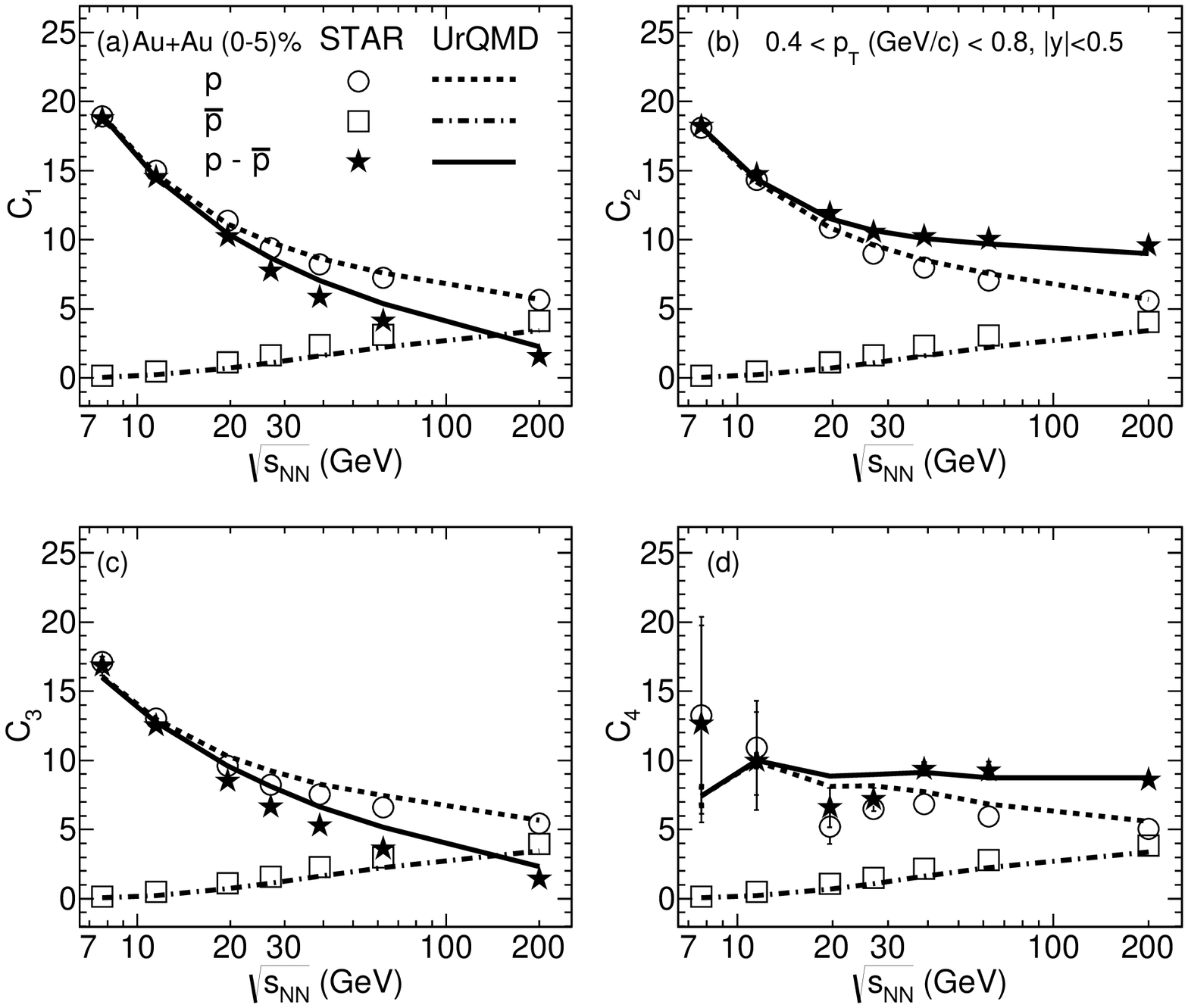}
\caption{Cumulants of proton, anti-proton and net-proton multiplicity
  distribution at midrapidity in 0-5\% central Au+Au collisions at RHIC BES program.
Data are compared to a transport model UrQMD calculation.}
\label{fig4}
\end{center}
\eef

A transport based model, Ultra Relativistic Quantum Molecular Dynamics
(UrQMD) model ~\cite{urqmd}  is expected to be effective in
explaining several bulk observables at lower colliding energies. 
It is based on a microscopic transport theory where the
phase space description of the reactions are important. It allows for the propagation
of all hadrons on classical trajectories in combination with stochastic binary 
scattering, color string formation and resonance decay. It incorporates baryon-baryon,
meson-baryon and meson-meson interactions, the collisional term includes more than 50 
baryon species and 45 meson species.
Figure~\ref{fig4} shows the comparison of the measured
cumulants for proton, anti-proton and net-proton  0-5\% central
distributions to the corresponding results from UrQMD calculations
obtained within the same acceptance as the experimental data~\cite{Luo:2013bmi}.
The UrQMD model results are reasonably close to the experimentally measured
$C_{1}$ and $C_{2}$ cumulants for proton, anti-proton and net-proton
distributions. However there is clear deviation from the
experimentally measured higher order cumulants of proton and
net-proton distributions for  $\sqrt{s_{NN}}$ = 19.6 and 27 GeV. The
anti-proton distributions follow reasonably well the UrQMD
expectations. Hence we conclude like for the case of Poisson and Binomial
baseline measures the UrQMD model results for higher order cumulants cannot explain
the measured values at $\sqrt{s_{NN}}$ = 19.6 and 27 GeV.

\subsection{Hadron Resonance Gas}

\bef
\begin{center}
\includegraphics[scale=0.40]{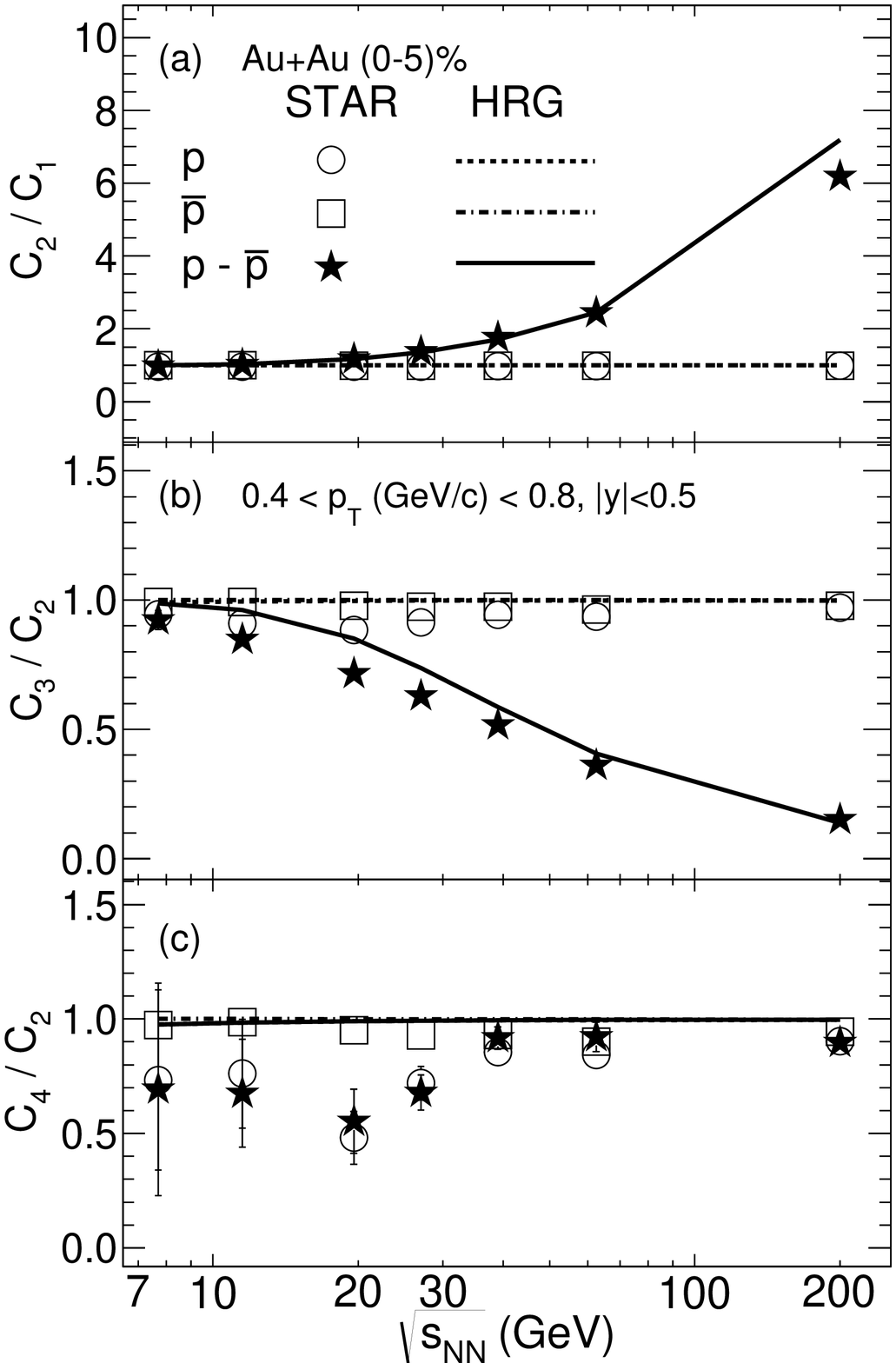}
\caption{Ratio of cumulants of proton, anti-proton and net-proton multiplicity
  distribution at midrapidity in 0-5\% central Au+Au collisions at RHIC BES program.
Data are compared to a hadron resonance gas model calculation~\cite{Garg:2013ata}.}
\label{fig5}
\end{center}
\eef

Thermal model calculations including hadrons and its resonances are
able to explain the yields of various hadrons produced in high energy
heavy-ion collisions~\cite{Cleymans:2005xv}. Here we compare the ratio of various measured
cumulants of the proton, anti-proton and net-proton distributions for
central 0-5\% Au+Au collisions at various beam energies to a hadron
resonance gas model calculation~\cite{Garg:2013ata}. The calculation is performed for
the same acceptance as the experimental data. The ratio of cumulants
are taken to cancel out the volume effect to first order in such a
model calculation. Figure~\ref{fig5} shows the comparison between data
and HRG values for $C_{2}/C_{1}$,  $C_{3}/C_{2}$, and
$C_{4}/C_{2}$. The $C_{2}/C_{1}$ ratio is reasonably well explained
for all the three (proton, anti-proton and net-proton) distributions
for all energies. $C_{3}/C_{2}$ and $C_{4}/C_{2}$ ratios from
anti-protons are reasonably well explained by HRG model. However the
measured proton and net-proton $C_{3}/C_{2}$ and $C_{4}/C_{2}$ ratios shows clear deviation
from HRG model values for $\sqrt{s_{NN}}$ = 19.6 and 27 GeV.

We find that all the four baseline considered for understanding the
experimental data on cumulants of measured proton, anti-proton and
net-proton distributions are consistent with the finding that the
0-5\% central Au+Au collision data at midrapidity for  $\sqrt{s_{NN}}$
= 19.6 and 27 GeV deviates from all them. This indicates data has
evidence for new physics process not included or expected from the
baseline measures considered.

\section{Understanding Independent Production Model}

In section II we have discussed why IP model explains the experimental
data on cumulants of net-proton distributions for  $\sqrt{s_{NN}}$ $<$
39 GeV. It is essentially due to the fact (as shown in the Fig.~\ref{fig1}) that
the cumulants of net-proton distributions are dominantly determined
from the cumulants of the corresponding proton distribution. However
the puzzle still remains that in-spite of significant correlated
production of proton and anti-protons at high energies, why the
measured cumulants follow so closely the IP model. This aspect is
discussed below using Heavy Ion Jet Interaction Generator (HIJING) model~\cite{hijing}. HIJING is an event generator
for heavy-ion collisions. It is a perturbative QCD inspired model which
produces multiple minijet partons, these later get transformed into string 
configurations and then fragment to hadrons.
In addition the role of baryon number conservation on such an IP model
as discussed in Ref~\cite{Bzdak:2012an}  is also discussed.

\subsection{Rapidity and Energy Dependence}
\bef
\begin{center}
\includegraphics[scale=0.40]{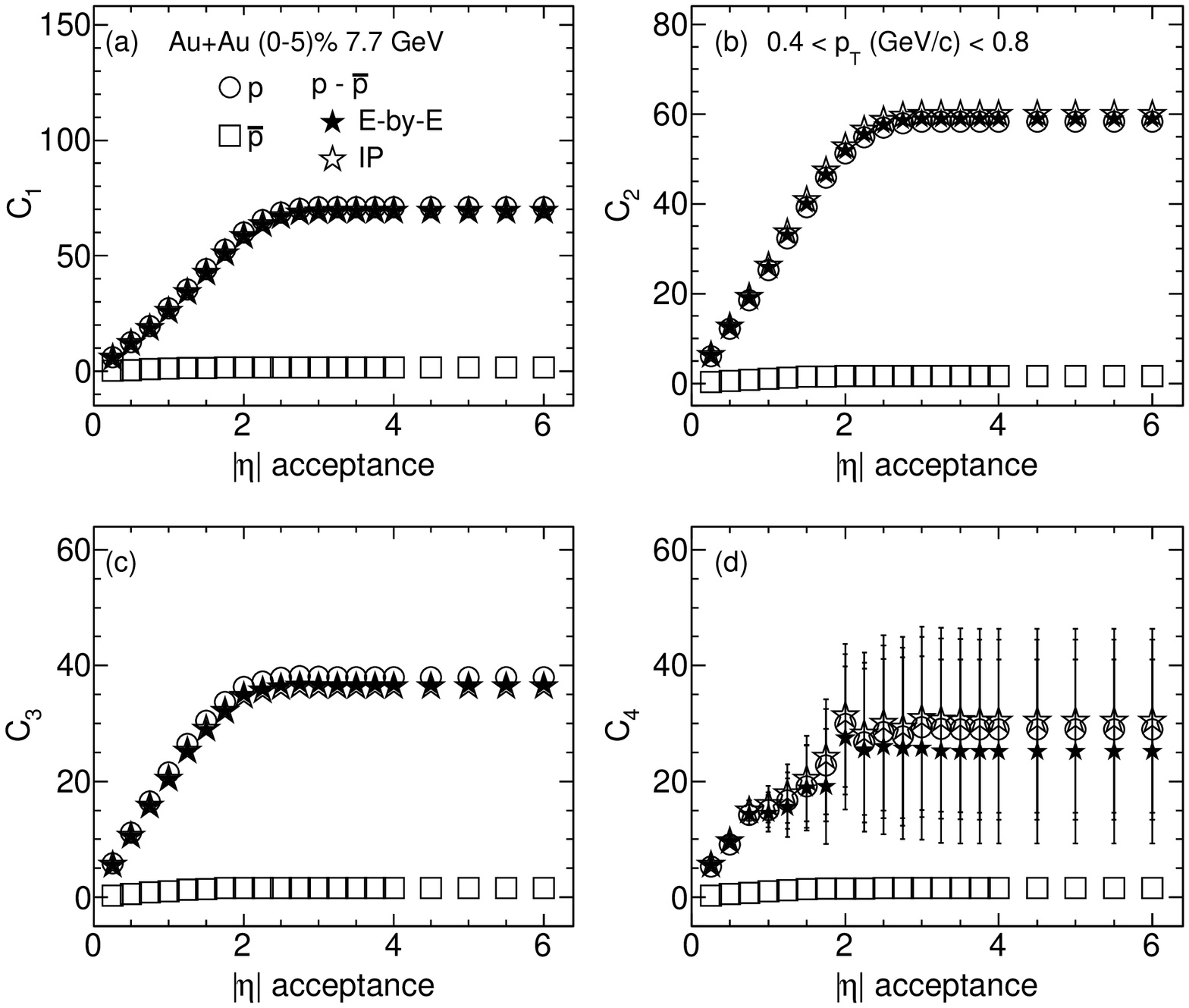}
\includegraphics[scale=0.40]{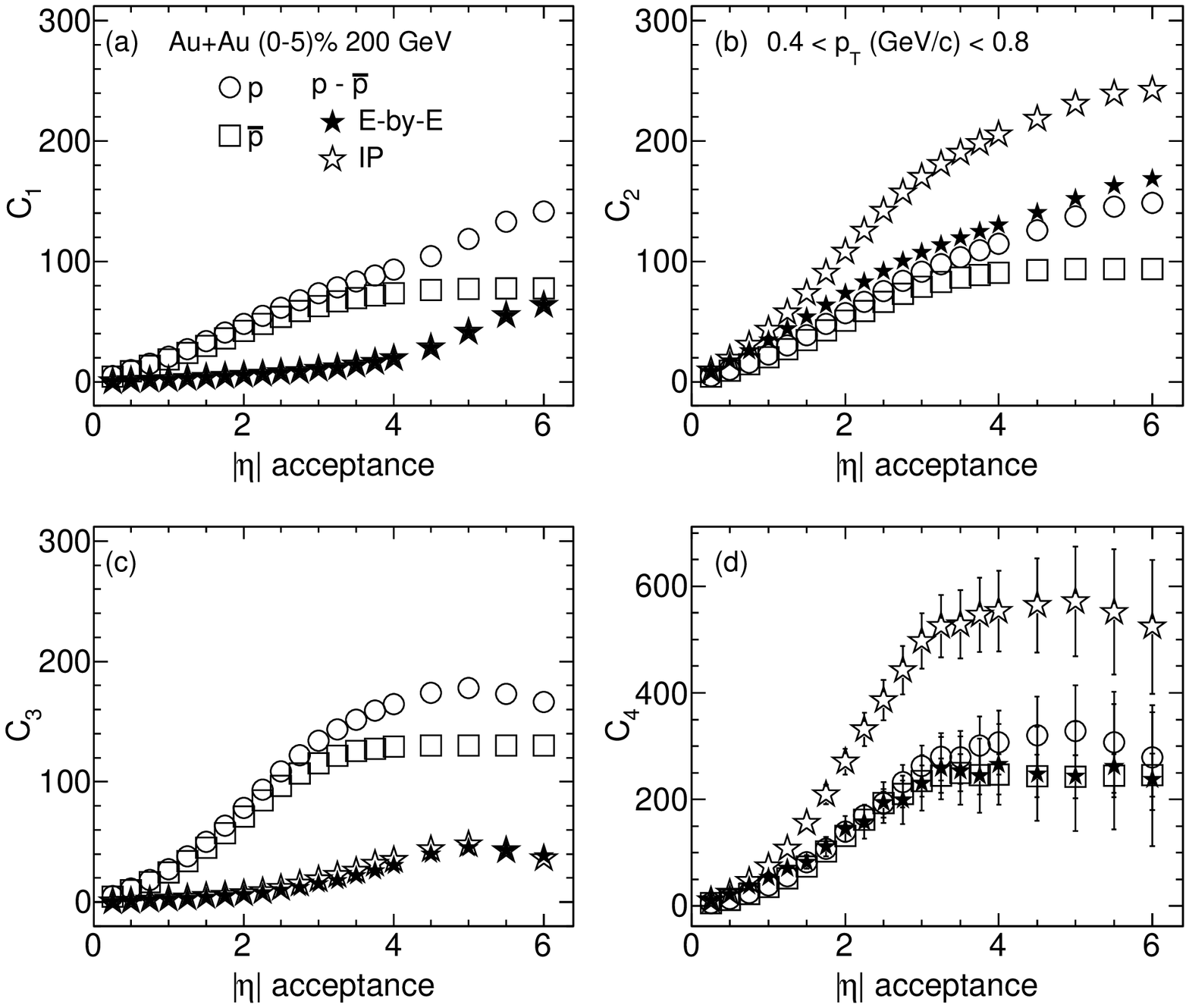}
\caption{Cumulants of proton, anti-proton and net-proton (E-by-E) multiplicity
  distribution as a function of pseudorapidity acceptance in 0-5\% central Au+Au collisions from
  HIJING at $\sqrt{s_{NN}}$ = 7.7 (top panel) and 200 GeV (bottom panel). Also shown is the
  expectation for net-proton from Independent Production (IP) model. The
error bars shown are statistical. The large statistical errors at
larger $\eta$ is due to higher values of $\sigma$ of the distribution.}
\label{fig5}
\end{center}
\eef

Figure~\ref{fig5} shows the cumulants of proton, anti-proton and
event-by-event (E-by-E) net-proton distribution from HIJING model in
Au+Au collisions for 0-5\% central collisions as a function of
pseudorapidity ($\eta$) acceptance. The results are shown for two
extreme beam energies (7.7 and 200 GeV) at RHIC for which data has
been collected in Au+Au collisions. All cumulants increases with
$\eta$ acceptance upto the corresponding beam rapidity and then
the values saturates for both $\sqrt{s_{NN}}$ = 7.7 and 200 GeV. There is hardly
any anti-protons produced at 7.7 GeV collisions, hence the net-proton
cumulants are dominated by the corresponding cumulants of the proton
distribution. Therefore the IP model expectation very closely follows the
net-proton and proton cumulant values. There by confirming our
expectation that IP model should not be considered as a baseline for
the lower beam energies.  In contrast we see interesting $\eta$
dependence for the Au+Au collisions simulated in HIJING at 200 GeV. 
A considerable amount of anti-protons are produced and we find the net-proton
$C_1$ and $C_3$ closely agree with IP expectation for the full $\eta$
range studied. However we see clear deviation of net-proton $C_2$ and
$C_4$ from the corresponding IP expectation for $\eta$ $>$ 0.5.
This study suggests that the agreement between data and corresponding
IP result is because of the $\eta$ acceptance of the measurement and
larger acceptance would perhaps have shown the deviations. The
deviations are as expected from the  breaking of the correlations due
to the proton and anti-proton pair production in IP construction at  
high beam energies. 

\subsection{THERMINATOR}
\bef
\begin{center}
\includegraphics[scale=0.40]{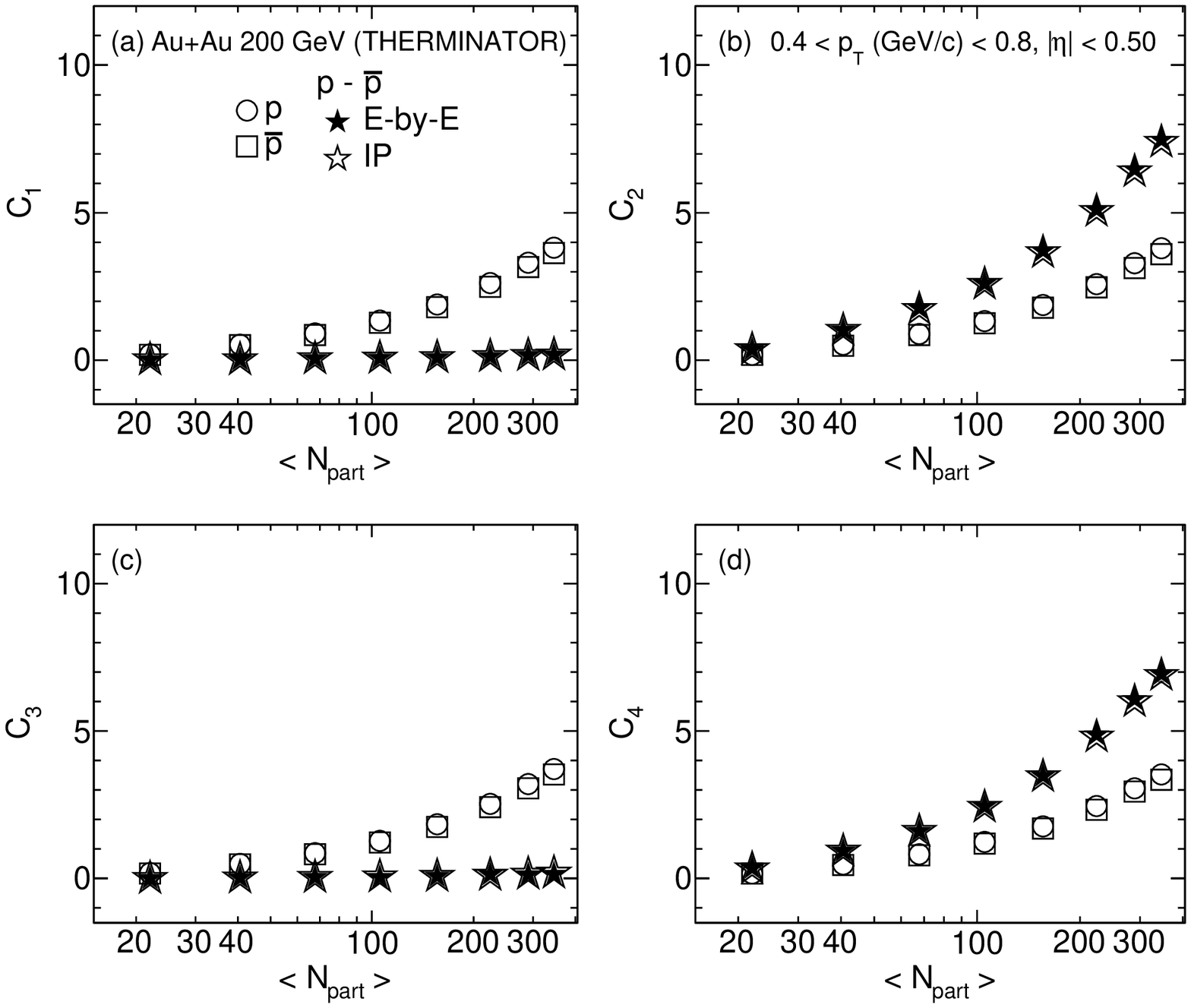}
\includegraphics[scale=0.40]{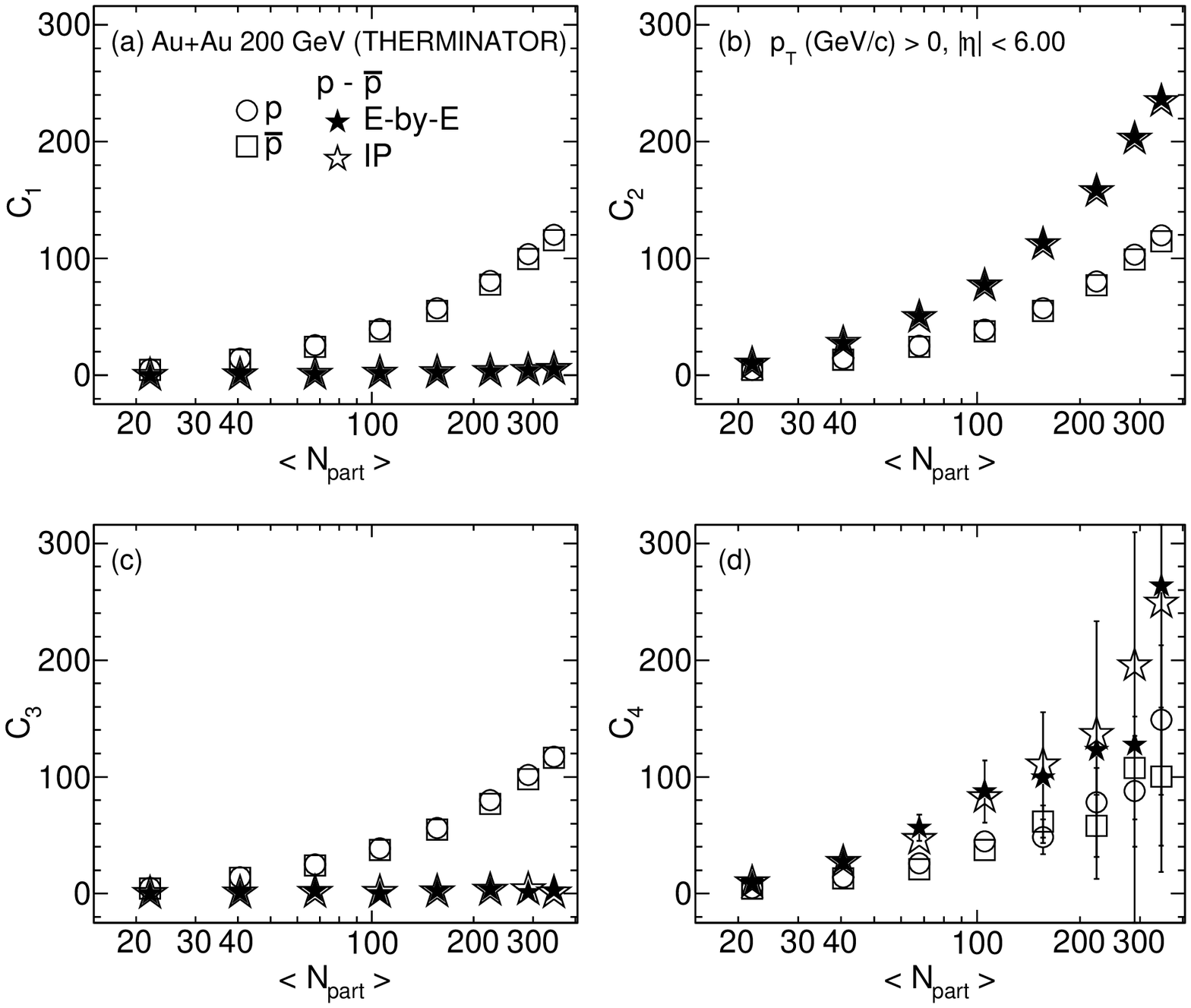}
\caption{Cumulants of proton, anti-proton and net-proton multiplicity
  distribution as a function of number of participating nucleons in Au+Au collisions from
  THERMINATOR~\cite{Kisiel:2005hn} at $\sqrt{s_{NN}}$ = 200 GeV. Also shown is the
  expectation for net-proton from Independent Production (IP) model.
  The top panel is for the $p_{T}$ range from 0.4 to 0.8 GeV/$c$ and
  $\mid \eta \mid$ $<$ 0.5, while the bottom panel is for $p_{T}$ $>$
  0 and $\mid \eta \mid$ $<$ 6.0.}
\label{fig6}
\end{center}
\eef

In this subsection we investigate another possibility of agreement
between IP expectation and data at higher beam energies. We consider
THERMINATOR event generator~\cite{Kisiel:2005hn}, which produces particles assuming
thermal equilibrium, grand canonical ensemble and ideal hydrodynamics.
 
Figure~\ref{fig6} shows the cumulants of proton, anti-proton and
net-proton distribution from THERMINATOR model simulated
for Au+Au collisions at $\sqrt{s_{NN}}$ = 200 GeV as a function of average
number of participating nucleons ($\langle N_{part} \rangle$). The
results shown for proton, anti-proton and net-proton
distributions are for same acceptance as the experimental data and also
for the full acceptance in transverse momentum and rapidity. The
net-proton cumulants are compared to corresponding IP
expectations. The IP expectations follows closely the net-proton
cumulant results for both the cases and all the collision centrality
studied. Since this feature is very different to that seen for HIJING
(Fig.~\ref{fig5}), it suggests that such an agreement with IP could be
due to the system being in thermal equilibrium and following a
statistical picture of a grand canonical ensemble. This study brings
in an alternate physics argument towards the reason for agreement of
IP expectation and the experimental data. 

\subsection{Baryon Number Conservation}

The net-baryon number probability distribution for both baryon and
anti-baryon distributions following Poisson distribution and with
baryon number conservation is given as~\cite{Bzdak:2012an} :

\begin{eqnarray}
P_{B}(n) &=&\left( \frac{p_{B}}{p_{\bar{B}}}\right) ^{n/2}\left( \frac{%
1-p_{B}}{1-p_{\bar{B}}}\right) ^{(B-n)/2}  \label{PB1finalG} \\
&&\times \frac{I_{n}\left( 2z\sqrt{p_{B}p_{\bar{B}}}\right) I_{B-n}\left( 2z%
\sqrt{(1-p_{B})(1-p_{\bar{B}})}\right) }{I_{B}(2z)},  \nonumber
\end{eqnarray}%

where $z=\sqrt{\left\langle N_{B}\right\rangle \left\langle N_{\bar{B}} \right\rangle}$, $\langle N_{B} \rangle$ is the average number of
      baryons, $\langle N_{\bar{B}} \rangle$  is the average number
      of anti-baryons and $I_{n}$ is the modified Bessel function of
      first kind. The analogues equation for net-proton is obtained
      from Eqn.1 by~\cite{Bzdak:2012an}
\begin{equation}
p_{B}=\frac{\langle n_{B}\rangle }{\langle N_{B}\rangle }\rightarrow \frac{\langle n_{p}\rangle }{\langle N_{B}\rangle },
\end{equation}%
where $\langle n_{p}\rangle $ is the mean number of observed protons
and analogously for anti-protons.

\bef
\begin{center}
\includegraphics[scale=0.40]{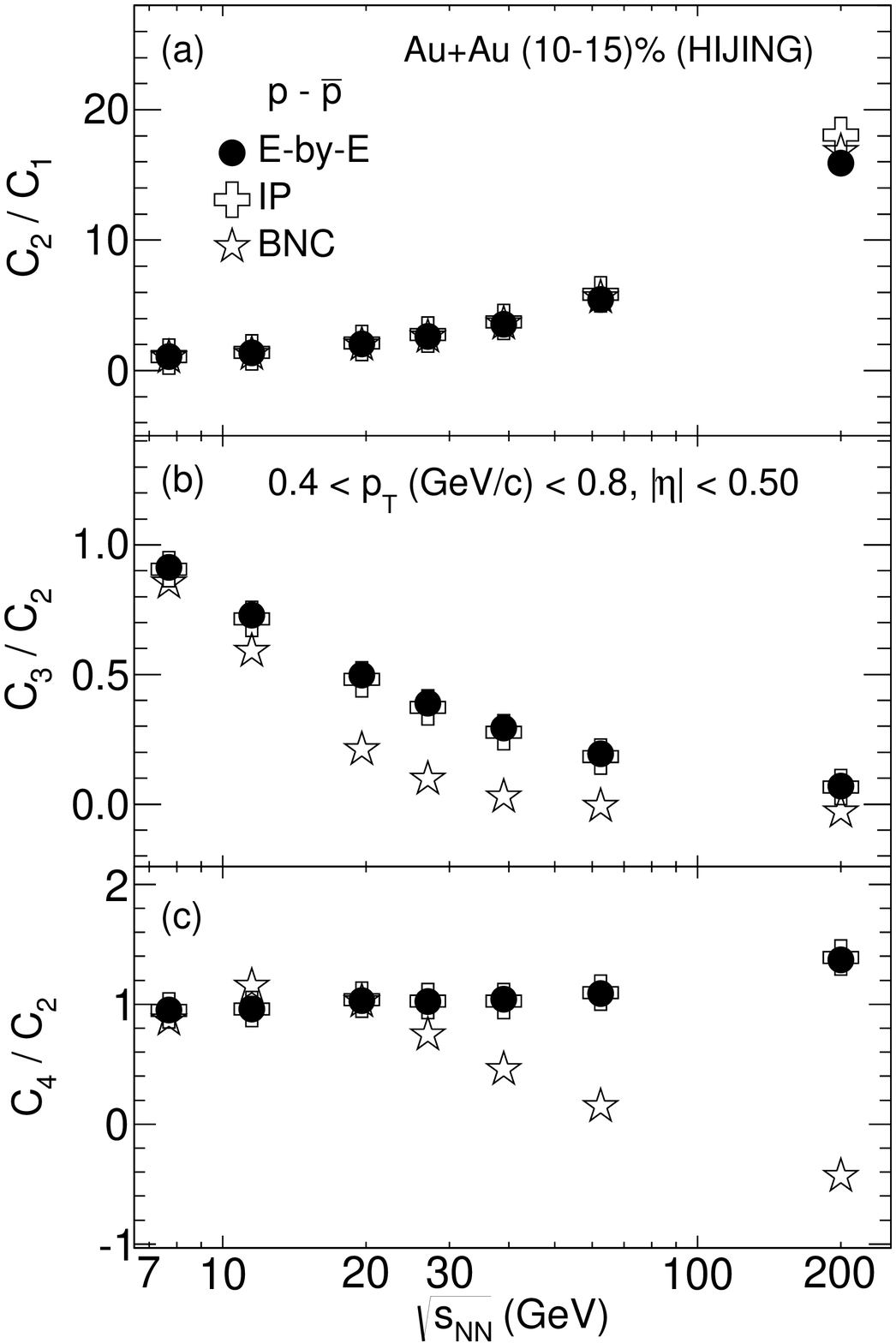}
\caption{Ratio of cumulants of net-proton multiplicity distribution as
  a function of beam energy in Au+Au collisions from HIJING. Also shown are the
  expectations for net-proton cumulant ratios from corresponding Independent Production
  approach with (BNC) and without (IP) baryon number conservation incorporated. }
\label{fig7}
\end{center}
\eef
Since it involves the knowledge of average number of total baryons and
anti-baryons produced in the collisions, we tested the comparison of
IP with and without baryon number conservation (BNC) in HIJING model.
Figure~\ref{fig7} shows the ratio of cumulants of net-proton
distributions in 10-15\% collision centrality for Au+Au collisions in
HIJING for various beam energies. This result is compared to
expectation from IP approach without baryon number conservation and
with baryon number conservation (denoted as BNC). While not much
difference is observed for the three cases for the ratio
$C_{2}/C_{1}$, differences between IP and BNC starts to show for
higher order ratios of $C_{3}/C_{2}$ and $C_{4}/C_{2}$ at higher beam
energies. This also emphasizes the need to properly construct the IP
approach considering the effect of baryon number conservation while
comparing to the experimental data. 

We have discussed in this section that IP approach needs to
carefully used as a baseline for comparison to data. Applicability of
IP approach at lower beam energies where shape of the net-proton
distribution is dominated by those from protons only is
questionable. At higher beam energies similarity between IP and data could be a
coincidence due to small acceptance. The importance of implementing
the baryon number conservation effect is emphasized. Further we find
the IP approach is also sensitive to the mechanism of particle production, as
one based on thermal model and ideal hydrodynamics as in THERMINATOR gives a different
conclusion to that based on HIJING.

\section{Summary}

In summary, we have discussed several possible baseline measures for understanding
the net-proton cumulants measured in STAR experiment at RHIC. The data
tends to show a non-monotonic variation in the higher order ratios of cumulants as a
function of beam energy. The baseline measures considered are those where the
underlying proton and anti-proton distributions are Poisson,
binomial, those generated from a transport model and a hadron
resonance gas model. Results from all baseline measures studied indicate that higher order
cumulants for 0-5\% Au+Au collision data deviates from them at
$\sqrt{s_{NN}}$ = 19.6 and 27 GeV. This indicates presence of new
physics, which could be related to a critical point in the strong
interaction phase diagram. A high statistics measurement at lower beam
energies as planned in the second phase of the RHIC beam energy scan
program and a detailed comparison to QCD calculations with critical
point will be able to settle the physics picture. 

We have discussed several drawbacks of comparing the data to a simple IP
approach. At lower beam energies the comparison is not valid as the
net-proton distribution are dominated by the shape of the proton
distribution only. This arises due to small yields of anti-protons at the
lower beam energies. Further at higher beam energies the agreement
between data and IP could be mere coincidence due to the acceptance
range used in the measurement. This aspect we have demonstrated using
a simulation study based on HIJING model. We have also pointed out
through the HIJING simulations that such IP approach also needs to
incorporate baryon number conservation effects. This is true for all
baseline measurements. Finally we have also
shown that the agreement between IP and the data could also
depend on the mechanism of particle production. In an event generator
like THERMINATOR that assumes particle production from a thermalized source shows no
difference between the model cumulant results and the corresponding IP result.

\noindent{\bf Acknowledgments}\\
BM is supported by the DST SwarnaJayanti project fellowship.

\end{document}